\documentclass[aps,prb,reprint,showpacs]{revtex4-1}
\usepackage{amssymb,amsmath,graphicx,gensymb}

\begin{document}

\title{Spin imbalance in hybrid superconducting structures with spin-active interfaces}

\author{Oleksii Shevtsov}
\email{shevtsov@chalmers.se}
\author{Tomas L\"{o}fwander}
\email{tomas.lofwander@chalmers.se}

\affiliation{Department of Microtechnology and Nanoscience -- MC2, Chalmers University of Technology, SE-412 96 G\"{o}teborg, Sweden}

\date{\today}


\begin{abstract}
We consider a heterostructure consisting of a normal metal and a superconductor separated by a spin-active interface. At finite bias voltages, spin-filtering and spin-mixing effects at the interface allow for an induced magnetization (spin imbalance) on the superconducting side of the junction, which relaxes to zero in the bulk. Such interfaces are also known to host a pair of in-gap Andreev bound states which were recently observed experimentally. We show that these states are responsible for the dominant contribution to the induced spin imbalance close to the interface. Motivated by recent experiments on spin-charge density separation in superconducting aluminum wires, we propose an alternative way to observe spin imbalance without applying an external magnetic field. We also suggest that the peculiar dependence of the spin imbalance on the applied bias voltage permits an indirect bound state spectroscopy. 
\end{abstract}

\maketitle

\section{Introduction}
Non-equilibrium phenomena in superconductors have attracted much attention since the pioneering works on charge imbalance by Clarke and co-workers.\cite{Clarke_PRL1972,Tinkham_Clarke_PRL1972,Tinkham_PRB1972,Chi_Clarke_PRB_1979} They found that an excess charge brought into a superconductor by tunneling electrons reduces the Cooper pair density close to the interface because of the charge neutrality constraint. This leads to a non-vanishing resistance of this part of the superconductor. The theoretical picture proposed to explain this effect \cite{Tinkham_Clarke_PRL1972} was based on imbalance between the number of electron-like and hole-like quasiparticles in the superconductor when the bias was higher than the superconducting gap. 

Each electron tunneling into the superconductor also brings along its spin moment. Therefore, if the number of injected electrons is different for opposite spin projections (e.g. by using a ferromagnet instead of a normal metal, or by applying a magnetic field) it is possible to induce a non-equilibrium magnetization, or spin imbalance, together with the charge imbalance at the superconducting side of the interface. In a normal metal, charge and spin of an electron are bound together. The nature of Bogoliubov quasiparticles in a superconductor is more complicated. Indeed, recent experiments\cite{Quay_NatPhys2013,Hubler_PRL2012,Wolf_PRB2013} have demonstrated spin and charge density separation,\cite{Kivelson_PRB1990} a situation when charge imbalance and spin imbalance relax away from the interface on different length scales. We note that in these experiments the orbital pair-breaking effect of an external magnetic field was needed to observe spin-charge density separation.

Here, we propose an alternative way to observe spin imbalance, which does not require a magnetic field. Our idea relies on the possibility of fabricating spin-active interfaces.\cite{MRS_PRB1988,Tokuyasu_PRB1988,Fogelstrom_PRB2000,ZLS_PRB2004,Cottet_PRB2005,Bobkova_Bobkov_2007,Kalenkov_Zaikin_PRB2007,Cottet_PRB2009,Linder_PRL2009,Linder_PRB2010,Bobkova_Bobkov_2011,Asano_PRB2012} One can imagine such interface as a magnetic layer with spin-dependent transmission amplitude and phase (via Larmor precession around the intrinsic magnetic moment of the layer). A superconductor coated with a spin-active layer hosts a pair of interface bound Andreev states, whose properties are controlled by parameters of the interface.\cite{Fogelstrom_PRB2000,ZLS_PRB2004} They have been observed in recent tunneling experiments on nanoscale superconductor-ferromagnet junctions.\cite{Hubler_PRL2012ABS} We show in this paper that these states give a dominant contribution to the spin imbalance effect near the interface and comment on the possibility of measuring this effect experimentally. 

The paper is organized as follows. In Sec.~\ref{Sec-Model} we describe the theoretical model of the spin-active interface, make a short introduction to the quasiclassical Green's function method,\cite{Larkin_Ovchinnikov_1969,Eschrig_PRB2000,Eschrig_PRB2009} and explain technical details of calculations. In Sec.~\ref{Sec-Results} we present the main results of the paper and discuss their relation to the recent experiments. Sec.~\ref{Sec-Summary} summarizes our findings and concludes the paper.

\section{Theoretical model}\label{Sec-Model}
\subsection{Spin-active interface}
\begin{center}
\begin{figure}[t]
\includegraphics[width=0.45\textwidth]{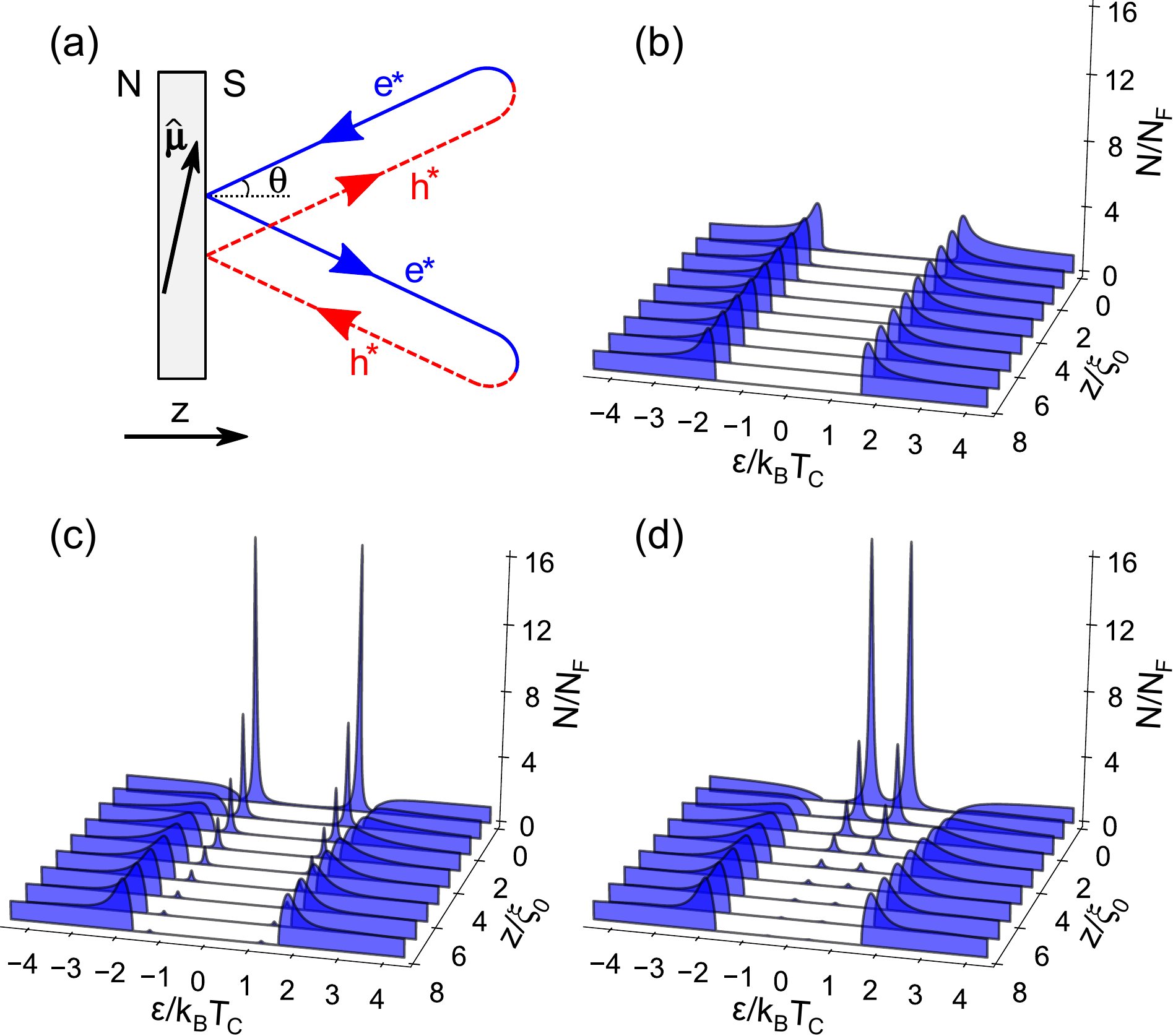}
\caption{(color online). (a) A normal metal (N) -- superconductor (S) junction with a spin-active interface is characterized by an intrinsic magnetic moment along $\hat{\mu}$. The closed trajectory indicates formation of Andreev surface bound states. (b)-(d) Local density of states in S as function of distance from the interface for spin mixing angles $\vartheta=0$, $0.49\pi$, and $0.83\pi$. The interface transparencies are $D_{\uparrow}=D_{\downarrow}=0.06$. Here, $\theta$ is the incidence angle, $\xi_0$ is the superconducting coherence length, $T_C$ the critical temperature, and $N_F$ the density of states at the Fermi energy in the normal state.} \label{Fig1}
\end{figure}
\end{center}
Consider a junction between a normal metal (N) at $z<0$, and a superconductor (S) at $z>0$, with a spin-active interface at $z=0$, as in Fig.~\ref{Fig1}(a). Assuming a smooth specularly reflecting interface invariant in the transversal direction, we may consider spatial dependence along the longitudinal z-axis only. A simple model of such an interface can be quantified by the following scattering matrix\cite{ZLS_PRB2004} connecting incoming and reflected electrons in the normal state,
\begin{align}
&\mathcal{S}=
\begin{pmatrix}
S_d & iS_{nd} \\
iS_{nd} & S_d
\end{pmatrix},\:
S_d=\left[R+\rho(\hat{\mu}\cdot\boldsymbol\sigma)\right]e^{i(\hat{\mu}\cdot\boldsymbol\sigma)\vartheta/2},
\notag\\
& S_{nd}=\left[D+\delta(\hat{\mu}\cdot\boldsymbol\sigma)\right]e^{i(\hat{\mu}\cdot\boldsymbol\sigma)\vartheta/2},
\label{SM}
\end{align}
with reflection coefficients $R=(\sqrt{R_{\uparrow}}+\sqrt{R_{\downarrow}})/2$, $\rho=(\sqrt{R_{\uparrow}}-\sqrt{R_{\downarrow}})/2$, and transmission coefficients $D=(\sqrt{D_{\uparrow}}+\sqrt{D_{\downarrow}})/2$, $\delta=(\sqrt{D_{\uparrow}}-\sqrt{D_{\downarrow}})/2$. They fulfill $R_{\uparrow,\downarrow}+D_{\uparrow,\downarrow}=1$.
Note that $\mathcal{S}$ in Eq.(\ref{SM}) is a $2\times2$ matrix in the ``left-right'' space, i.e. $\left[\mathcal{S}\right]_{11}$ refers to the reflection from N to N, while $\left[\mathcal{S}\right]_{12}$ describes the transmission from N to S. In addition, each element of $\mathcal{S}$ is itself a $2\times2$ matrix in spin space, where $\boldsymbol\sigma$ is a vector of spin Pauli matrices. We have written explicitly the scattering matrix for particles, $\mathcal{S}$. The corresponding scattering matrix for holes is given by $\mathcal{S}_{h}=\tilde{\mathcal{S}}^{\dagger}$, see Eq.(\ref{TildeOp}) and Ref.~\onlinecite{Eschrig_PRB2009}.

\begin{center}
\begin{figure}[t]
\includegraphics[width=0.45\textwidth]{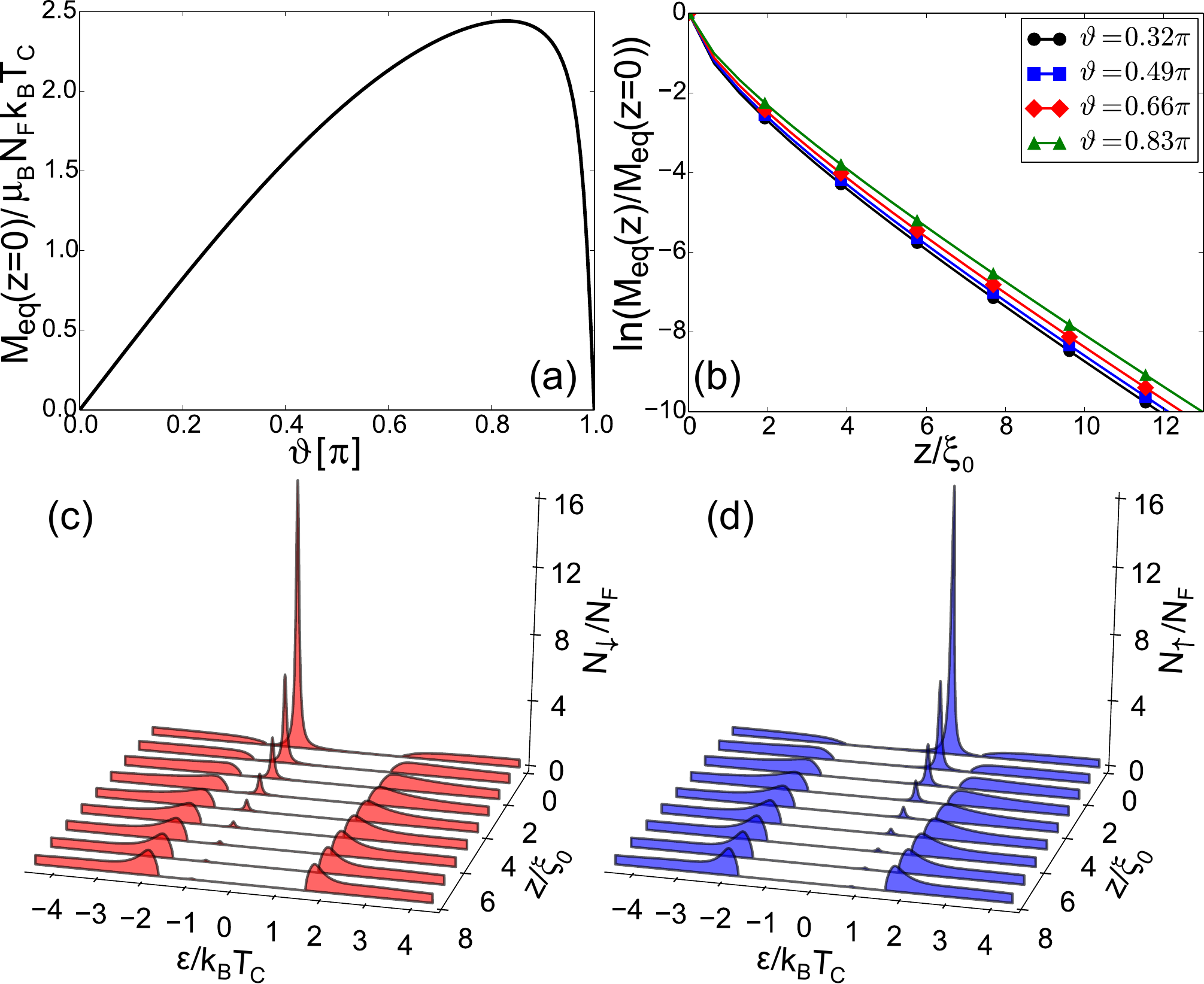}
\caption{(color online). (a) Interface value $M_{eq}(0)$ as function of spin-mixing angle. (b) Semi-log plot of $M_{eq}(z)/M_{eq}(0)$ as function of distance from the interface. (c)-(d) Spin-down and spin-up local density of states $N_{\downarrow,\uparrow}(\epsilon,z)$ as function of distance for $\vartheta=0.66\pi$. $D_{\uparrow}=D_{\downarrow}=0.06$, and $T=0.01T_C$.} \label{Fig2}
\end{figure}
\end{center}

For an impenetrable wall ($R_{\uparrow}=R_{\downarrow}$=1), reflections are accompanied by spin-dependent phase shifts through the spin-mixing angle $\vartheta$.\cite{Tokuyasu_PRB1988,Fogelstrom_PRB2000,ZLS_PRB2004} This leads to formation of surface bound states, trapped between the impenetrable wall and the bulk of the superconductor by the superconducting gap $\Delta$ in the spectrum. A Bohr-Sommerfeld quantization rule can be set up\cite{Lofwander_2001} by considering the closed loop in Fig.~\ref{Fig1}(a). A spin-mixing phase $\pm\vartheta/2$ is picked up during reflection at the interface, where the signs correspond to spin-up and spin-down states. An energy ($\epsilon$) dependent phase shift  $-\gamma(\epsilon)\mp\chi$ is picked up during Andreev reflection, where the signs correspond to electron-hole and hole-electron conversion processes. Here, $\gamma(\epsilon)=\arccos(\epsilon/\Delta)$, and $\Delta$ and $\chi$ are the magnitude and phase of the superconducting order parameter, respectively.\footnote{The superconducting phase $\chi$ drops out (it can also be gauged away), but is kept here for generality.} The quantization condition becomes $\vartheta-2\gamma(\epsilon)=2n\pi$, where $n$ is an integer. The resulting surface states appear at energies $\epsilon_{ABS}=\pm\Delta\cos(\vartheta/2)$.\cite{Fogelstrom_PRB2000,ZLS_PRB2004} The wave functions of the surface states decay into the bulk of the superconductor at a characteristic length $\xi_{ABS}(\theta)\simeq\hbar v_F\cos\theta/\sqrt{\Delta^2-\epsilon_{ABS}^2}$, where $\theta$ is the angle between the quasiparticle trajectory and the z-axis, see Fig.~\ref{Fig1}(a), and $v_F$ is the quasiparticle velocity at the Fermi surface in the normal state. This length scale can be very long if the bound state is close to the gap edge $\epsilon_{ABS}\lesssim\Delta$. However, after averaging over all angles, as in the local density of states, Fig.~\ref{Fig1}(c)-(d), the bound state peak still decays\cite{Metalidis_PRB2010} at a short distance of the order of the superconducting coherence length $\xi_0=\hbar v_F/2\pi k_BT_C$, where $T_C$ is the critical temperature. For a tunnel barrier ($D_{\uparrow,\downarrow}\ll 1$), the surface states broaden into resonances of width $\sim D_{\uparrow,\downarrow}\Delta$. As confirmed experimentally,\cite{Hubler_PRL2012ABS} the positive- and negative-energy resonance peaks correspond to quasiparticle states with opposite spin projections, see Fig.~\ref{Fig2}(c)-(d). We note that when $\vartheta=0$, there are no bound states at the interface, see Fig.~\ref{Fig1}(b).
\subsection{Quasiclassical Green's function}
For the calculations we utilize the quasiclassical Green's function formalism\cite{Larkin_Ovchinnikov_1969,Eschrig_PRB2000,Eschrig_PRB2009} and the goal is to calculate the function $\check{g}(\epsilon,\mathbf{p}_{F},\mathbf{r},t)$. Here $\epsilon$ is the quasiparticle energy, $\mathbf{p}_{F}$ is the quasiparticle momentum on the Fermi surface, $\mathbf{r}$ is the spatial coordinate, and $t$ is the time. Below we will omit function arguments for brevity. This function has a $2\times2$ matrix structure in Keldysh space denoted by "check", 
\begin{equation}
\label{gKeldMatr}
\check{g}=
\begin{pmatrix}
\hat{g}^{R} & \hat{g}^{K} \\
0 & \hat{g}^{A} 
\end{pmatrix},
\end{equation}
and a $2\times2$ matrix structure in Nambu (or particle-hole) space denoted by "hat",
\begin{gather}
\hat{g}^{R,A}=
\begin{pmatrix}
g^{R,A} & f^{R,A} \\
\tilde{f}^{R,A} & \tilde{g}^{R,A}
\end{pmatrix},\;
\hat{g}^{K}=
\begin{pmatrix}
g^{K} & f^{K} \\
-\tilde{f}^{K} & -\tilde{g}^{K}
\end{pmatrix},\label{g_RAK}
\end{gather}
where $g^{R,A,K}$, $f^{R,A,K}$, etc. are $2\times2$ spin matrices. It satisfies the quasiclassical Eilenberger equation\cite{Eilenberger1968}
\begin{equation}
\label{EilEqn}
[\epsilon\hat{\tau}_3 \check{1} - \check{h}, \check{g}]_{\otimes} +i\hbar \mathbf{v}_{F}\cdot\boldsymbol{\nabla} \check{g}
= \check{0},
\end{equation}
where the self-energy matrix $\check{h}$ is parametrized as
\begin{gather}
\check{h}=
\begin{pmatrix}
\hat{h}^{R} & \hat{h}^{K} \\
0 & \hat{h}^{A} 
\end{pmatrix},\;\notag\\
\hat{h}^{R,A}=
\begin{pmatrix}
\Sigma^{R,A} & \Delta^{R,A} \\
\tilde{\Delta}^{R,A} & \tilde{\Sigma}^{R,A}
\end{pmatrix},\;
\hat{h}^{K}=
\begin{pmatrix}
\Sigma^{K} & \Delta^{K} \\
-\tilde{\Delta}^{K} & -\tilde{\Sigma}^{K}
\end{pmatrix},\label{SC_Eqs_general}
\end{gather}
and $\Sigma^{R,A,K}$, $\Delta^{R,A,K}$, etc. are spin matrices. We introduce the ``tilde''-operation defined by
\begin{align}
\tilde{Y}(\varepsilon,\mathbf{p}_{\mathrm{F}},\mathbf{r},t)=Y(-\varepsilon^{\ast},-\mathbf{p}_{\mathrm{F}},\mathbf{r},t)^{\ast}.\label{TildeOp}
\end{align}
where $\varepsilon=\epsilon$ for the Keldysh components and $\varepsilon=\epsilon\pm i0^{+}$ for retarded and advanced components, respectively. The matrices $\hat{\tau}_3$ and $\check{1}$ are third Pauli matrix in Nambu space and unity matrix in Keldysh space. Equation (\ref{EilEqn}) has to be supplemented by the normalization condition
\begin{equation}
\check{g}\otimes\check{g}=-\pi^2\check{1},\label{gNormCond}
\end{equation} 
where the $\otimes$-product is defined by
\begin{equation}
\check{A}\otimes\check{B}(\epsilon,t)=e^{i\hbar(\partial^{A}_{\epsilon}\partial^{B}_t-\partial^{A}_{t}\partial^{B}_{\epsilon})/2}\check{A}(\epsilon,t)\check{B}(\epsilon,t).
\end{equation}

We employ the Riccati parametrization\cite{Schopohl_Maki_PRB1995,Schopohl_1998,Eschrig_PRB2000,Eschrig_PRB2009} for the elements of Eq.(\ref{gKeldMatr}). Then Eq.(\ref{EilEqn}) and Eq.(\ref{gNormCond}) transform into a system of equations, which can be solved efficiently either analytically or numerically. On the other hand, to solve Eq.(\ref{EilEqn}) and Eq.(\ref{gNormCond}) near the interface we have to specify appropriate boundary conditions. This is a non-trivial question because the interface modeled by a sharp boundary cannot be described quasiclassicaly. Therefore one has to derive effective boundary conditions. This problem was solved by several authors\cite{Zaitsev_1984,MRS_PRB1988,Eschrig_PRB2000,Fogelstrom_PRB2000,ZLS_PRB2004,Eschrig_PRB2009} for the Eilenberger equation and others\cite{Kuprianov_Lukichev_1988,Nazarov_1999,Cottet_BCs_2009} for the Usadel equation\cite{Usadel_1970} (which is obtained as a diffusive limit of the Eilenberger equation). In our work we use the boundary condictions derived in Ref.~\onlinecite{Eschrig_PRB2009} which take the scattering matrix Eq.(\ref{SM}) as an input. These equations are rather lengthy and are not rewritten here. We note that in this paper we study stationary non-equilibrium and the time coordinate $t$ drops out. The $\otimes$-product then reduces to simple matrix multiplication.

Finally, in general, Eq.(\ref{EilEqn}) has to be solved self-consistently together with the corresponding self-consistency equations for the self-energies Eq.(\ref{SC_Eqs_general}). In particular, the order parameter of an s-wave singlet superconductor $\Delta^{R}_0(\mathbf{r})=i\sigma_2\Delta_0(\mathbf{r})$ reads
\begin{align}
\Delta_0(\mathbf{r})=-\frac{i\lambda N_{F}}{8\pi}\int_{-\epsilon_c}^{\epsilon_c} d\epsilon\int\frac{d\Omega_{\mathbf{p}_{F}}}{4\pi}\,\mathrm{Tr}\left[i\sigma_2f^{K}(\epsilon,\mathbf{p}_{F},\mathbf{r})\right],\label{s-waveOP}
\end{align}
where $\lambda<0$ is the electron-phonon coupling constant and $\epsilon_c$ is the high-energy cut-off of the order of the Debye frequency. $\Delta_0(\mathbf{r})$ is a scalar complex-valued function.

As soon as the Green's function is known one can calculate various physical observables,\cite{Serene_Rainer_PhysRep1983} such as spin imbalance,
\begin{align}
\mathbf{M}(\mathbf{r})&= 2\mu_{B}^2N_{F}\mathbf{B}(\mathbf{r})\notag\\
&+\frac{i\mu_{B}N_{F}}{8\pi}\!\int\! d\epsilon\int\!\frac{d\Omega_{\mathbf{p}_{F}}}{4\pi}\,\mathrm{Tr}\left[\hat{\boldsymbol{\alpha}}\hat{g}^{K}(\epsilon,\mathbf{p}_{F},\mathbf{r})\right],\label{SI}
\end{align}
and local density of states,
\begin{align}
N(\epsilon,\mathbf{r})=-\frac{N_{F}}{2\pi}\mathrm{Im}\left\{\int\!\frac{d\Omega_{\mathbf{p}_{F}}}{4\pi}\mathrm{Tr}\left[\hat{\tau}_3\hat{g}^{R}(\epsilon,\mathbf{p}_{F},\mathbf{r})\right]\right\}.
\end{align}
Here $\hat{\boldsymbol{\alpha}}=\mathrm{diag}(\boldsymbol{\sigma},\boldsymbol{\sigma}^{\ast})$ is a block-diagonal matrix in Nambu space, $\mathbf{B}(\mathbf{r})$ is an external magnetic field, $\mu_B$ is the Bohr magneton, $e$ is the electron charge, and $N_F$ is the density of states at the Fermi level in the normal state.
\begin{center}
\begin{figure*}[t]
\includegraphics[width=0.92\textwidth]{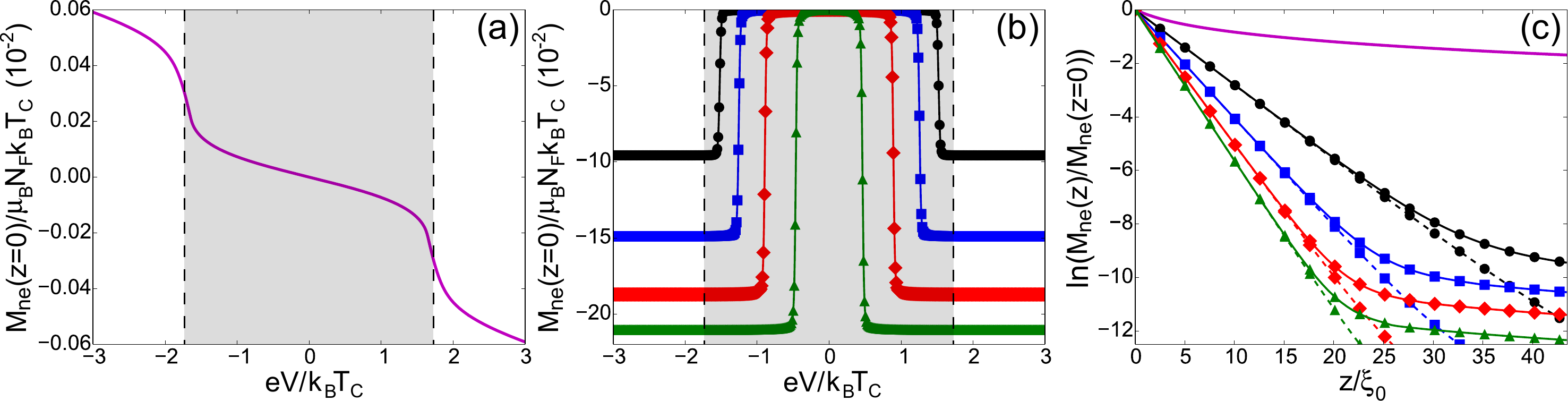}
\caption{(color online). Non-equilibrium part of spin imbalance. (a) Interface value as a function of bias voltage for $\vartheta=0$, $D_{\uparrow}=0.06$, and $D_{\downarrow}=0.02$. (b) Interface value for $\vartheta=0.32\pi$ (black circles), $\vartheta=0.49\pi$ (blue rectangles), $\vartheta=0.66\pi$ (red diamonds), $\vartheta=0.83\pi$ (green triangles), and $D_{\uparrow}=D_{\downarrow}=0.06$. (c) Semi-log plot of $M_{ne}(z)/M_{ne}(0)$ for $eV=3.2k_BT_C$ (solid lines) and $eV=1.7k_BT_C$ (dashed lines). The grey rectangle in (a)-(b) depicts the subgap region. Temperature $T=0.01T_C$.}\label{Fig3}
\end{figure*}
\end{center}
\subsection{Details of the calculation}
We assume that a finite bias voltage $V$ is applied across the NS junction, Fig.~\ref{Fig1}(a). It is convenient to split the Keldysh Green's function into spectral and anomalous
parts,\cite{Eschrig_PRB2000,Eschrig_PRB2009}
\begin{align}
\hat{g}^K=\left[\hat{g}^R-\hat{g}^A\right]\tanh\frac{{\epsilon}}{2k_BT}
+\hat{g}^a.\label{gK_sp_a}
\end{align}
To avoid confusion we stress that the term ``anomalous'' in this context describes the deviation from equilibrium. It should not be confused with the off-diagonal Green's function in Nambu space, $f^{R}$ in Eq.(\ref{g_RAK}), which describes superconducting electron-hole coherence, and is also sometimes called ``anomalous'' in the literature. Here $\hat{g}^a$ describes pure non-equilibrium effects due to applied bias voltage $V$ and has both diagonal and off-diagonal elements in Nambu space. Then the spin imbalance Eq.~(\ref{SI}) is expressed as
\begin{equation}
\mathbf{M}(z)=\mathbf{M}_{eq}(z)+\mathbf{M}_{ne}(z).
\end{equation}
The first term corresponds to the spectral part of $\hat{g}^K$. It exists in equilibrium and is sometimes called inverse proximity effect. \cite{Bergeret_EPL2004,Bergeret_PRB2004,Bergeret_PRB2005,Xia_PRL2009,Grein_PRB2013} The second term, related to the anomalous propagator $\hat{g}^a$ in Eq.(\ref{gK_sp_a}), is a true non-equilibrium contribution and it depends explicitly on applied bias voltage. 

There are several mechanisms responsible for spin relaxation,\cite{ZFD_RMP2004} among which are scattering against magnetic impurities or presence of spin-orbit coupling in combination with momentum scattering by e.g. scalar impurities. In this work we focus on the simplest mechanism, namely scattering by magnetic impurities characterized by a spin-flip length $l_{sf}=v_{F}\tau_{sf}$, see Ref.~\onlinecite{Bennemann_Ketterson_2008}, 
\begin{align}
\check{h}_{sf}(\epsilon,z)=\frac{\hbar}{2\pi\tau_{sf}}\int\frac{d\Omega_{\mathbf{p}_{F}}}{4\pi}(\hat{\tau}_3\check{1})\check{g}(\epsilon,\mathbf{p}_{F},z)(\hat{\tau}_3\check{1}),
\end{align}
where $\tau_{sf}$ is the spin-flip time.
The presence of a small fraction of magnetic impurities can significantly reduce the order parameter.\cite{Abrikosov_Gorkov_1961,Phillips_PRL1963,Bennemann_Ketterson_2008} We consider the case $l_{sf}\gg\xi_0$, for which the pair breaking effect is small. For the calculations we use $l_{sf}\approx300\xi_0$ and compute the bulk impurity self-energy and the bulk order parameter self-consistently. We obtain $\Delta\approx1.776k_BT_C$, which is a bit higher than the usual BCS value because of the presence of magnetic impurities: the critical temperature $T_C$ decreases faster than the order parameter as a function of magnetic impurities concentration.\cite{Abrikosov_Gorkov_1961,Bennemann_Ketterson_2008} We note that the order parameter is real in our case since we neglect, for simplicity, superfluid momentum in the superconducting region as it has a small effect on spin imbalance. 

The results presented below are for the experimentally relevant tunneling limit, $D_{\uparrow,\downarrow}\ll 1$. In this case, surface states have a well-defined energy. In the tunneling limit, and for small spin-mixing angles $\vartheta$, the order parameter is only marginally suppressed near the interface and self-consistency of self-energies may be neglected when computing spin imbalance. Below we focus on such a non-self consistent calculation (also for arbitrary $\vartheta$) of interface properties and comment in the end on the effects of self-consistency. Finally, since there is a single spin quantization axis in the problem given by the interface moment $\hat{\mu}$, the spin imbalance is parallel to it, $\mathbf{M}(z)=M(z)\hat{\mu}$.
\begin{center}
\begin{figure*}[t!]
\includegraphics[width=0.92\textwidth]{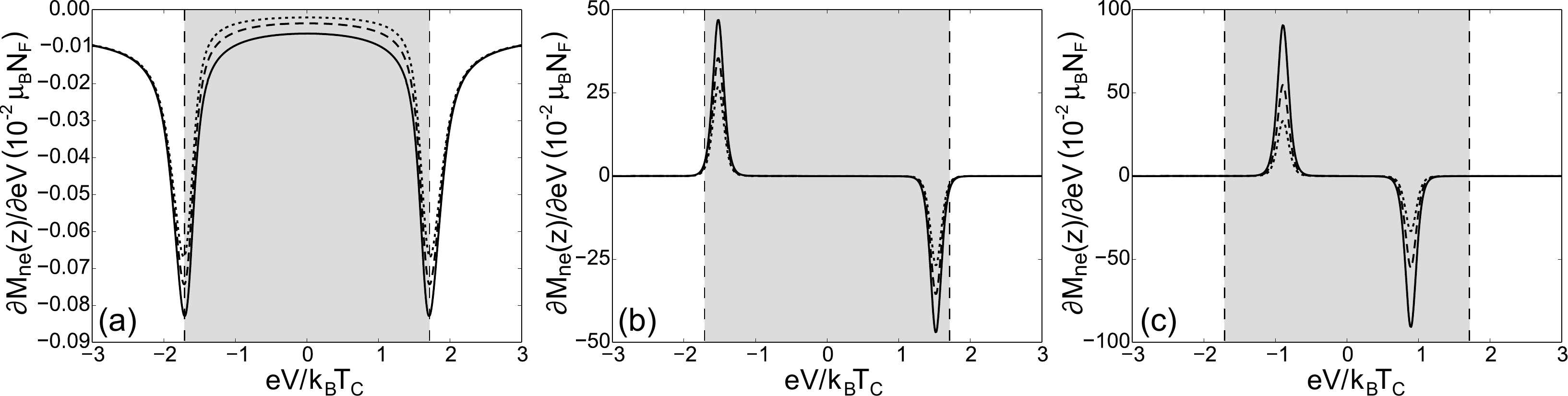}
\caption{Derivative of the anomalous part of spin imbalance with respect to bias voltage for: (a) $\vartheta=0$, $D_{\uparrow}=0.06$, and $D_{\downarrow}=0.02$; (b) $\vartheta=0.32\pi$ and $D_{\uparrow}=D_{\downarrow}=0.06$ ; (c) $\vartheta=0.66\pi$ and $D_{\uparrow}=D_{\downarrow}=0.06$. Solid, dashed and dotted lines correspond to $z=0,\xi_0$, and $2\xi_0$, respectively. The grey rectangle depicts the subgap region. Temperature $T=0.05T_C$.}\label{Fig4}
\end{figure*}
\end{center}
\section{Results and discussion}\label{Sec-Results}
\subsection{Spectral part of spin imbalance: inverse proximity effect}
We start by discussing the spectral part of spin imbalance, see Fig.~\ref{Fig2}. By definition,
\begin{align}
M_{eq}(z)\!=\!\frac{\mu_B}{2}\!\int\! d\epsilon\left[N_{\uparrow}(\epsilon,z)-N_{\downarrow}(\epsilon,z)\right]\tanh\frac{\epsilon}{2k_BT},
\end{align}
where $N_{\uparrow,\downarrow}(\epsilon,z)$ are spin-resolved local densities of states. The magnitude of $M_{eq}(z)$ is determined by the weight of the Andreev states in the total density of states, see Fig.~\ref{Fig2}(c)-(d). In Fig.~\ref{Fig2}(a) we plot the interface value as function of spin-mixing angle. The decrease of $M_{eq}(0)$ for large values of $\vartheta$ is due to overlap (in energy space) of the bound state peaks. One can show that $M_{eq}(z)=\sin\vartheta f(z,\vartheta)$, see Ref.~\onlinecite{Grein_PRB2013}, and vanishes at $\vartheta=0$ and $\pi$. Note that $M_{eq}(z)$ is not purely sinusoidal because of the function $f(z,\vartheta)$, as can be seen in the figure. In Fig.~\ref{Fig2}(b) we show how the equilibrium part of spin imbalance decays away from the interface. It turns out that contributions to $M_{eq}(z)$ from bound states and from continuum states quickly cancel each other as we move into the bulk. Therefore, the inverse proximity effect decays on the short coherence length scale $\xi_0$ independently of $\vartheta$.\cite{Bergeret_EPL2004,Bergeret_PRB2004,Bergeret_PRB2005,Grein_PRB2013} We emphasize that the spectral contribution $M_{eq}(z)$ exists at zero bias and is a consequence of the interface-induced difference between the spin-resolved densities of states, and no direct quasiparticle injection is needed. Therefore the decay of this contribution is governed by a healing length of a superconductor, which is $\xi_0$.
\subsection{Non-equilibrium part of spin imbalance}
Let us now consider the non-equilibrium part\footnote{For the calculations of the non-equilibrium part $M_{ne}(z)$ we use a tunneling cone $D_{\uparrow,\downarrow}(\theta)=D_{\uparrow,\downarrow}(0)e^{-\beta\sin^2\theta}$,\cite{Wolf_2012} which is an experimentally relevant model. We used $\beta=80$, which corresponds to a $14\degree$ wide tunneling cone. The transmission probabilities $D_{\uparrow,\downarrow}$ in the text are the normal incidence values $D_{\uparrow,\downarrow}(0)$. Note that it is not necessary to use the tunneling cone model for $M_{eq}(z)$ since it is proportional to reflection coefficients, which are close to unity for all $\theta$.} of spin imbalance, $M_{ne}(z)$, see Fig.~\ref{Fig3}.
There are two main contributions: spin-filtering ($\vartheta=0$, $D_{\uparrow}\neq D_{\downarrow}$) and spin-mixing ($\vartheta\neq 0$, $D_{\uparrow}=D_{\downarrow}$). In Fig.~\ref{Fig3}(a)-(b) we plot the interface values, $M_{ne}(0)$, as function of bias voltage for these two components. They have different symmetries under $V\rightarrow -V$. The spin-filtering component is an odd function since positive and negative biases correspond to adding or withdrawing majority spins. The spin-mixing component is an even function because positive and negative biases correspond either to populating an Andreev surface state at positive energy with one spin projection or depopulating the corresponding negative energy state with the opposite spin projection. Furthermore, their voltage dependences are different. The spin-filtering component is due to injection into continuum states and depends on the size of the bias window and grows linearly at large bias, but is quenched in the sub gap region. The spin-mixing component, Fig.~\ref{Fig3}(b), consists of a sharp increase of spin imbalance for voltages corresponding to the energy of the surface state, but saturates quickly when the whole resonance lies in the bias window, since the difference between spin-resolved densities of states is small in the continuum $|eV|>\Delta$. For general parameters, the two components are superimposed (not shown), but we note that the spin-mixing component dominates at the interface because the bound state is completely spin polarized and its occupation leads to a large spin imbalance. We emphasize that the non-equilibrium part of spin imbalance is a result of a direct quasiparticle injection, $\hat{g}^a(\epsilon,\mathbf{p}_{F},z)\propto D_{\uparrow,\downarrow}$. Hence the difference in magnitude between the spin-filtering and spin-mixing contributions. The former is a result of tunneling into the continuum of states above the gap, while the latter is predominantly due to the sub-gap bound state resonance.

In Fig.~\ref{Fig3}(c) we show the spatial dependences of the spin-filtering and spin-mixing contributions. For the case of pure spin-filtering and $eV>\Delta$, we inject spin polarized quasiparticles into continuum states. This contribution relaxes through scattering against magnetic impurities and decays on the spin-flip length scale, $l_{sf}$ [slowly decaying, magenta line, in Fig.~\ref{Fig3}(c)]. For the case of pure spin-mixing and $\epsilon_{ABS}<eV<\Delta$, we populate only the Andreev bound state and $M_{ne}(z)$ decays, after averaging\cite{Note2} over the Fermi surface, on the length scale $\xi_{ABS}^{eff}=\hbar v_F/\sqrt{\Delta^2-\epsilon_{ABS}^2}$ (dashed lines) that can be long when the bound state is close to the gap edge (small $\vartheta$). When $eV>\Delta,\epsilon_{ABS}$, we also populate a fraction of continuum states. Then, close to the interface the spatial dependence is determined by the Andreev bound state, while for distances far enough that the bound state has decayed, the dominant contribution comes from the continuum with decay length $l_{sf}$ (solid lines).
\subsection{Relation to experiment}
Let us now discuss the implications of our results for experiments. In Ref.~\onlinecite{Wolf_PRB2013}, a non-local differential conductance of a NISIF structure ("I" stands for insulator and "F" for ferromagnet) $g_{nl}\propto dI_{det}/dV_{inj}$ in an external magnetic field was measured. Here, $I_{det}$ is the current at the detector electrode in response to an injection voltage $V_{inj}$. For analysing the data they used the tunnel model,\cite{Zhao_Hershfield_PRB1995}
\begin{align}
I_{det}=\frac{G_{det}}{eN_F}\left[(Q^{\ast}_{\uparrow}+Q^{\ast}_{\downarrow})+P_{det}(S_{\downarrow}-S_{\uparrow})\right],\label{Tunnel_model_Idet}
\end{align}
where $G_{det}$ is the normal state detector conductance, $P_{det}$ is the detector spin polarization, $Q^{\ast}_{\uparrow,\downarrow}$ are the spin-up/down contributions to charge imbalance,\cite{Hubler_PRL2012} and $S_{\uparrow,\downarrow}$ are the spin-up/down densities induced by a spin-polarized current.\cite{Hubler_PRL2012} In order to separate the spin and charge imbalance parts they used symmetries: the charge imbalance $Q^{\ast}=Q^{\ast}_{\uparrow}+Q^{\ast}_{\downarrow}$ is anti-symmetric with respect to $V_{inj}$, since for negative values electrons are injected into the system, while holes are injected for positive values. The spin imbalance, created by an external magnetic field, is symmetric with respect to $V_{inj}$. Therefore $I_{det}^{sym}(V_{inj})\propto (S_{\downarrow}-S_{\uparrow})$ and $g_{nl}^{asym}(V_{inj})\propto d(S_{\downarrow}-S_{\uparrow})/dV_{inj}$.
Note that the differential conductance has opposite symmetry to the current. At the same time the induced magnetization is $M=(|e|g/2m)(S_{\downarrow}-S_{\uparrow})$, where $g$ is the electron g-factor and $m$ is the electron effective mass. Therefore the non-local signal, within the model of Eq.(\ref{Tunnel_model_Idet}), is $g_{nl}^{asym}(V_{inj})\propto dM/dV_{inj}$.

In the experiment, the external magnetic field was crucial as the spin imbalance was created through spin polarization of the superconducting density of states (Zeeman effect).\cite{Tedrow_Meservey_PRL1971,Tedrow_Meservey_PhysRep1994} In our case the spin polarization comes from the interface-induced Andreev states. In another experiment, Ref.~\onlinecite{Quay_NatPhys2013}, they measured a non-local differential resistance in a FISIF structure. Again, an external magnetic field was crucial to observe spin imbalance because otherwise the only source of spin imbalance is the spin-filtering effect, which has the same symmetry as the charge imbalance and is much smaller.\cite{Quay_NatPhys2013} Finally, the (orbital) pair-breaking effect of the external magnetic field\cite{Maki_PTP1964} made the charge imbalance signal decay faster\cite{Hubler_PRB2010} than the spin imbalance in both experiments, a situation that was called spin-charge density separation.\cite{Kivelson_PRB1990,Quay_NatPhys2013}

In Fig.~\ref{Fig4}(a)-(c) we plot the derivative of $M_{ne}(z)$ with respect to bias voltage. Note that $M_{eq}(z)$ is independent of $V$ and is not relevant for these experiments. The spin-filtering part, Fig.~\ref{Fig4}(a), has the same symmetry with respect to $V$ as the non-local conductance due to charge imbalance (see Refs.~\onlinecite{Wolf_PRB2013,Quay_NatPhys2013}) and it cannot be separated from the latter by the symmetry arguments used above. That is why it was not observed in the experiment in Ref.~\onlinecite{Quay_NatPhys2013}. The derivative of the spin-mixing contribution, Fig.~\ref{Fig4}(b)-(c), resembles the non-local conductance due to spin imbalance in Ref.~\onlinecite{Wolf_PRB2013}. We note that  peaks observed experimentally occurred at voltages near the gap edges. In our case the peak positions as well as their decay length $\xi_{ABS}^{eff}$ are determined by the bound state energies $\pm\epsilon_{ABS}$. Therefore spin imbalance measurements can be used for bound state spectroscopy. Our analysis suggests that it is possible to observe the spin imbalance signal by doing analogous non-local measurements without applying an external magnetic field. We leave for future studies a quantification of spin-charge density separation in our setup, since it is necessary to compute the order parameter self-consistently to properly describe charge imbalance. For spin imbalance, self-consistency is not as crucial.

For highly transparent junctions, the width of the Andreev bound state is proportional to the barrier transparency while its weight in the total density of states is proportional to the reflection coefficient.\cite{ZLS_PRB2004} Thus, for the case of high transparency junctions, the resulting spin imbalance signal will be reduced and it will be difficult to assign a single decay length to the Andreev resonance states. We therefore conclude that it is desirable to work with spin-active tunnel junctions.

In case of a disordered sample, the mean free path $l$ reduces the superconducting coherence length,  $\xi_0\rightarrow\xi=\sqrt{l\xi_0/3}$.\cite{Kopnin_2009} In our model this means that the results we presented above hold but the length scale is reduced to $\xi$. In fully self-consistent calculations, the disorder broadens the bound states,\cite{Fogelstrom_PRB2000} but we believe our results to be still valid. Consequently, to test our predictions, clean samples give better spatial resolution of spin imbalance.

\section{Summary}\label{Sec-Summary}
In summary, we have computed spin imbalance in a normal metal--superconductor hybrid structure with a spin-active interface at finite bias voltage. The interface-induced Andreev bound states, existing at subgap energies, play a dominant role in creating the spin imbalance effect. For distances of the order of tens of superconducting coherence lengths away from the interface, spin imbalance relaxes with the characteristic length $\xi_{ABS}^{eff}$ set by the bound state. Currently used non-local conductance measurement techniques can in principle be used to observe this effect experimentally, as it possesses the same symmetry as the Zeeman-induced spin imbalance signal already observed in recent experiments, and is of opposite symmetry to the charge imbalance signal. The advantage of our setup is that it does not require an external magnetic field and that the characteristics of the spin imbalance are controlled by parameters of the interface, which can in principle be engineered.\cite{Khaire_PRL2010}

\begin{acknowledgments}
We acknowledge financial support from the Swedish Research Council.
\end{acknowledgments}

\bibliography{References_PRB}
\end{document}